\documentstyle[epsfig]{ipproc}
\def\eq#1{(\ref{#1})}
\def\be{\begin{equation}}
\def\ee{\end{equation}}
\begin{document}
\title{Coherent States in High-Energy Physics}

\author{C.S. Lam}
\address{Department of Physics, McGill University\\
3600 University St., Montreal, QC, Canada H3A 2T8}

\maketitle

\begin{abstract}
The amplitude for emitting $n$ bosons factorizes
into the product of $n$ single-boson emission amplitudes,
if the source is energetic and abelian. If it is energetic but 
{\it non-abelian}, the amplitude
is given by a sum of factorized 
{\it quasi-particle} amplitudes. A quasi-particle is made up of
an arbitrary number of bosons, but couples to the source like a single
one. 
Factorization is related to coherence, and it allows 
computation of subleading contributions not obtainable by usual
means. Its importance
is illustrated in two applications: 
to solve the baryon problem in large-$N_c$ QCD, and to
obtain a total cross section satisfying the Froissart bound.
\end{abstract}

\section{Introduction}
We found a quasi-particle state of gluons whose existence has eluded detection 
all these years. 
In this talk I will discuss how that comes  about,
and what use we can make of it.
A quasi-particle is made up of an arbitrary number of gluons, 
but it couples to
their high-energy source like a single one: as
a colour-octet object whose emission preserves helicity of the source.
Quasi-particles emerge naturally as a result of factorization and
coherence. They are present in all 
non-abelian theories, including the Yukawa theory of nucelons
and pions, and not just QCD. 

By a high-energy source,
I mean a source with large {\it total} energy. The source may be a highly
relativistic particle with a small mass, or a non-relativistic particle with
a very large mass. For simplicity, I shall refer to these two cases 
respectively as a relativistic source and a non-relativistic source.

By a non-abelian theory, I mean one
in which the
spin and/or the internal quantum numbers of the high-energy source
can be altered by the emission of bosons. 
In the case of pions emitted
from a massive non-relativistic nucleon, it is 
the spin and the isospin of the
nucleon that are affected by the emission. 
In the case of QCD it is the colour of the source.
But in the case of photons emitted
from a relativistic electron, neither the charge nor the helicity of
the electron is changed, so in this case the source is abelian.

After sketching the origin of the quasi-particle, and its connection with
factorization and cohernece, I will discuss two cases in which it makes its
presence felt. I believe these applications barely scratch the surface,
and the importance of these quasi-particles goes far beyond these two 
examples, but exactly in what way remains to be seen.

\begin{figure}[b!] 
\centerline{\epsfig{file=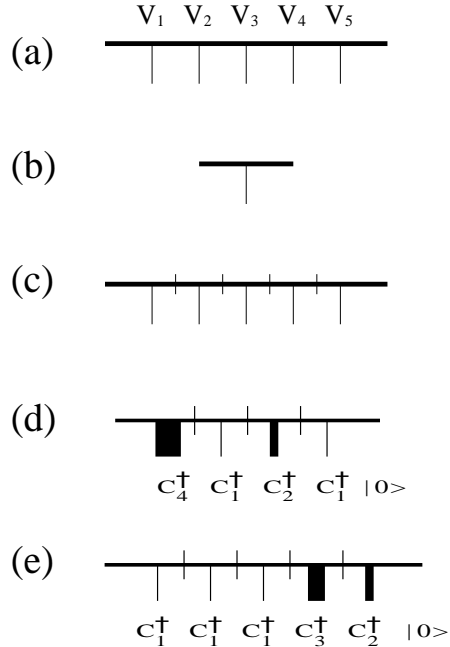,height=3.5in,width=3.5in}}
\vspace{10pt}
\caption{(a). A Feynman tree amplitude with vertex factors $V_i$; (b).
A one-boson amplitudes with on-shell sources; (c). The factorization of
a Bose-Einstein symmetrized amplitude into the product of one-boson
amplitudes; (d) and (e). Two separate factorization into quasi-particle
amplitudes.}
\label{fig1}
\end{figure}

\section{The Emergence of Quasi-Particles}
Consider the tree diagram Fig.~1(a), in which $n$ bosons 
are emitted from an energetic source via 
vertex factors $V_i$.  The source is assumed to be
energetic so that recoils suffered from the emissions can be ignored.
As a result, its transverse position $x_{\perp}$ 
is fixed, and so is the $x_\perp$ of every boson emerging 
from it. 
In the case of a relativistic source, we may
assume it to move parallel to the
$z$-axis near the speed of light, hence
its $x^-\equiv
x^0-x^3$ coordinate is fixed, thereby determining also
the $x^-$ coordinate of all
the emerging bosons. 
For a non-relativistic source,
its $x^3$ coordinate and that of the bosons are fixed. 
All in all,  three out of four coordinates of all the off-shell bosons
are identical. If the fourth
one  is also the same for all the bosons, a coherent state will
emerge. It turns out that  non-trivial inputs are required
to determine this fourth coordinate to achieve coherence.

It is necessary to invoke Bose-Einstein symmetry, which in  this case 
simply means summing up the $n!$ permuted tree amplitudes.
For abelian sources, the vertices $V_i$'s can be regarded 
simply as numbers. In that case we will show that every boson of
 the BE-symmetrized amplitude
has its momentum component $k^-=k^0-k^3=0$ if the source is
relativistic, and $k^0=0$ if the source is non-relativistic. 
This is the fourth coordinate
we are after for a coherent state.

The conclusion follows as a result of factorization.
After the summation, each boson
is allowed to be emitted anywhere along the 
tree, irrespective of the location of the others.
Hence the $n$-boson amplitude is factorized into a product of $n$
single-boson emission amplitudes. This is depicted in Fig.~1(c), where
a vertical cut on the tree indicates factorization.
For a relativistic source that is on-shell,  its momentum component
$p^-=p^0-p^3=0$, so by momentum conservation (see Fig.~1(b)) $k^-=0$
for every boson as claimed. For a non-relativistic source that is on-shell,
$p^0$ is fixed at its on-shell mass $M$, so by momentum conservation 
$k^0=0$. 
Note that neither conclusion is valid for the boson momenta
in a Feynman diagram like Fig.~1(a), where off-shell sources
are involved. In that case, uncertainty in energy prevents $p^-$
or $p^0$ to be fixed, so it is not true that $k^-$ or $k^0$ is zero.
Bose-Einstein symmetrization
and the resulting factorization are crucial to reach these conclusions.

Note also that this kind of coherent state is very different from
those encountered at low temperatures, where Bose-Einstein condensation
may occur. The coherent state we have is described by a mixture of
spatial and momentum coordinates, and it is not an energy eigenstate.
What is `cold' in the present context is the lack of recoil, instead
of the lack of thermal fluctuation.
Hence the physics outcome between the two are completely different as well.

For non-abelian sources
this simple factorization is no longer valid. The vertex factors
$V_i$ fail to commute with one another, so correction terms involving their commutators must be added \cite{LAM}. It turns out that
each of these correction terms is still factorizable,
but generally into products of {\it quasi-particle} amplitudes
instead of single-boson amplitudes.
In other words, it is the $k^-$ or the $k^0$ coordinates
of the quasi-particles as a whole that are zero, but not the individual
bosons within each quasi-particle.
As remarked before, a quasi-particle may consist of any number $k$ of
bosons, but instead of coupling to the source via the product of vertex
factors $V_1V_2\cdots V_k$, it does so via the nested commutator
$[V_1,[V_2,[\cdots,[V_{k-1},V_k]\cdots]]]$. In the case of QCD when 
$V_i$ are colour matrices, the nested commutator is given by 
a linear sum of
colour matrices, so the quasi-particle couples just like a colour-octet
object. Moreover, since each gluon making up the quasi-particle
does not flip the helicity of the source, neither will the quasi-particle.

Exactly how each correction term factorizes depends on the permutation.
I list here three examples for $n=8$: $[1|2|3|4|5|6|7|8],\ 
[8521|3|74|6],\ [1|2|3|864|75]$, 
in which a vertical bar indicates the position where
factorization takes place. The general rule is simply that
a bar should be put behind a number iff no number to its right is smaller
than it. The first example is identical to abelian factorization.
In the second example, the source emits a quasi-particle of four
gluons, one of two gluons, and two with one gluon each (a quasi-particle
with one gluon is just a gluon). In the third example, there are three
quasi-particles with one gluon each, one with three gluons, and one with
two gluons. The last two examples involve nested commutators so they
will not be present for abelian sources. In fact, other than the first
example, no other permutation can contribute in the case
of an abelian source
for exactly the same reason.

Letting $Q$ denote a quasi-particle, the general structure of each
factorized amplitudes is therefore of the form
\begin{equation}
[Q|Q|\cdots|Q],
\label{qqq}\end{equation}
where the different quasi-particles $Q$ appearing in this equation may
consist of different number of gluons.

\section{Composite Source} 
Suppose the source is made up of $N$ constituents, each capable of emitting a boson via the vertex $V_i=\psi^\dagger t_i\psi$, where $\psi$ and
$\psi^\dagger$ are the annihilation and creation operators 
for the constituents. For example, the source may be a nucleus with
$N$ nucleons, or a nucleon with $N$ quarks.
Being a one-body operator,
the matrix elements of $V_i$ is expected to be of order $N$.
Being a $k$-body operator, the matrix element of a product of $k$ $V_i$'s 
is expected to be of order $N^k$.
In contrast, the nested commutator of $k$ $V_i$'s is of the form $\psi^\dagger T\psi$,
with $T$ given by the nested commutator of the $t_i$'s,
so it is still a one-body operator and
 its matrix element is proportional to $N$.  If the an $n$-boson amplitude
in \eq{qqq} contains
$p$ quasi-particles, then that term is of order $N^p$, 
with $p\le N$, and not
$N^n$ that each Feynman tree diagram is expected to have.
Except for the identical permutation whose amplitude factorizes
completely as in
 $[1|2|3|\cdots|n]$, so that $p=n$, all the others have $p<n$
and hence contribute subdominantly when $N\gg 1$. The smaller $p$
is the less it contributes. 
If for some reason all the terms with $n\ge p\ge p_0+1$ vanish, then
the amplitude is of order $N^{p_0}$. It can still be computed easity
from the quasi-particle amplitudes with $p\le p_0$, but it
is extremely difficult to calculate it directly from Feynman tree diagrams,
especially when $p^0\ll n$. To do so
we must compute each Feynman diagram
to the subleading order $N^{p_0}$ before a finite sum can be obtained
upon summation, a highly non-trivial task.

Such a behaviour indeed happens 
in the process $\pi+{\cal N}\to (n-1)\pi+{\cal N}$, calculated in
large-$N$ QCD \cite{LARGEN}. In that case, the nucleon ${\cal N}$ consists
of $N$ quarks. Its mass is of order $N$ so it is a non-relativistic
energetic source. The effective
Yukawa interaction $t=g(\vec\sigma\cdot\vec k)
(\vec \tau\cdot\vec\pi(\vec k))$ is non-abelian because it flips the spin
and the isospin of the nucleon. Each Feynman tree amplitude is of order
$N^n/\sqrt{N}^n$ because it consists of $n$ vertices and the 
propagators are of order 1. A normalization factor $1/\sqrt{N}$
per pion is put in as usual \cite{LARGEN}. The amplitude
is huge for every $n>0$ in the limit $N\gg 1$. In this strong-coupling
limit one might think that very little could be said
about the reaction, and certainly
the Feynman-diagram description is useless 
even when loops are included. Yet the phenomenology of baryons
in large-$N$ QCD is very successful in describing nature \cite{BARYONS}.
What happens is that when the $n!$ permuted diagrams are summed up, a 
tremendous amount of
cancellation takes place, so that the final $n$-pion amplitude is of
order $N^{1-n/2}$ rather than $N^{n/2}$ of the individual diagrams.
The total pion-nucleon amplitudes now become weak 
for $n>2$, so we can understand
why loops are not needed and why phenomenology can be successful. In order
to prove this cancellation in a brute-force way, each diagram must be
calculated down to $(n-1)$ subleading orders, for at the end 
everything else
above it will be cancelled in the sum. This is quite an impossible task
for large $n$, and this is where the advantage of the factorized formula
\eq{qqq} shows up \cite{LL1}. 
If the number $p$ of quasi-particle amplitudes in 
\eq{qqq} is larger than 1, then it vanishes 
because of energy conservation. In that case one of these
$p$ factorized components must consist of only outgoing pions, which
 violates energy conservation since the nucleons are on-shell. 
As a result we are left with only terms with $p=1$,
whose matrix element is $N/\sqrt{N}^n=N^{1-n/2}$, as claimed.

\section{Damping Explosive Total Cross Sections}
Total cross section can be obtained from the forward elastic amplitude
via the optical theorem. This amplitude is difficult to compute
even assuming the coupling constant
$\alpha_s$ to be small, for at high cm energy complicated loop
diagrams must be included. This is so because
each time we add a loop to the diagram, we
add an extra factor of $\alpha_s$ but the loop integration may also
produce an extra $\ln s$. Thus a diagram of order $2n+2$
may give a contribution proportional to $\alpha_s(\alpha_s\ln s)^n$,
which is of order $\alpha_s$ if $\xi\equiv\alpha_s\ln s=O(1)$.
This is why diagrams of all orders must be included.

Computing multi-loop diagram is a difficult task which can be
accomplished only with suitable approximations. In the {\it leading-log
approximation}, which keeps only the lowest power of $\alpha_s$ while
keeping the variable $\xi$ fixed, the total cross section
so computed is proportional to 
$\alpha_s^2\exp(4\alpha_s\ln s \ln 2N_c/\pi)$. This is the famous
BFKL formula \cite{BFKL}. With $\alpha_s\simeq 0.19$, for example,
the cross section
according to this formula grows with energy like $s^{0.5}$. At this rate
the size of a proton becomes ten times the size of the Uranium nucleus
at LHC energy, and one hundred times its size at $10^{20}$ eV, the highest
energy cosmic-ray reaching earth. For better or for worse, this alarming
growth is not realized. In fact, Froissart bound forbids the total
cross section to grow faster than $\ln^2s$ at asymptotic energies. 
The theoretical challenge then is how to produce sufficient amount of 
corrections in QCD to satisfy the Froissart
bound. Since the BFKL computation already includes all the
important contributions in the leading-log approximations, 
{\it viz.,} all terms of order $\alpha_s^2$ when $\xi$
is kept fixed, clearly subleading terms of order $\alpha^m$ 
with $m\ge 3$ are needed for the Froissart bound.
As explained in the last section,
the factorization formula \eq{qqq} is capable of extracting
subleading terms $N^p$ for $p<n$ in that case. Similarly,
\eq{qqq} can be used to extract subleading terms
$\alpha_s^m$ with $m\ge 3$ \cite{DKL}. The result is shown in Fig.~1(b),
where the thick vertical lines represent quasi-particles, and the 
thin vertical cuts on the two horizontal lines represent factorization.
It can be shown that an amplitude with $p$ vertical quasi-particle
lines is of order $\alpha_s^p$; this is analogous to the situation
of the last section in which an amplitude with $p$ quasi-partilces
is proportional to $N^p$.

\begin{figure}[b!] 
\centerline{\epsfig{file=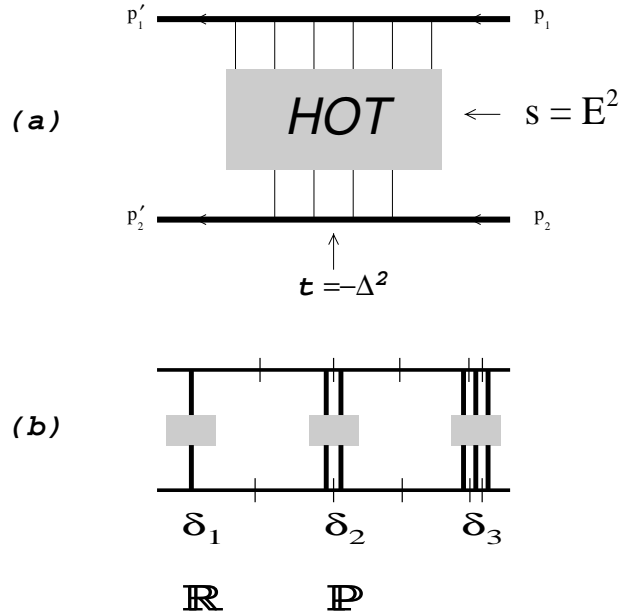,height=3.5in,width=3.5in}}
\vspace{10pt}
\caption{(a). A two-particle collision diagram showing a hot central
region and a `cold' peripheral region; (b). Factorization of
the scattering amplitude into quasi-particle amplitudes. An
amplitude with $p$ quasi-particle exchanges can be shown to be of
order $\alpha_s^p$ if $\alpha)s\ln s$ is of order 1.}
\label{fig2}
\end{figure}

For $\alpha_s\ll 1$ and $\xi=O(1)$, the dominant scattering
amplitude comes from diagrams with $p=1$ (alone), indicated in Fig.~2(b)
by $\delta_1$. Remebering that each quasi-particle carries an octet
colour, we conclude that the dominant amplitude is a colour-octet
amplitude (in the $t$-channel), whose magnitide is $O(\alpha_s)$.
It has been known long ago \cite{REGGEON} that the dominant amplitude
is obtained by the exchange of a colour-octet Reggeon, so from these
two equivalent descriptions
we can identify the quasi-particle with the Reggeon. What has thus been
achieved here is an algebraic characterization of the Reggeon,
as the colour-octet object obtained 
through factorization and coherence, rather than a pole in the angular
momentum plane as is usually defined.

A quasi-particle in QCD is not the same as a gluon, but a quasi-particle
in QED is identical to a photon because all the nested commutators vanish.
This distinction is ultimately 
the reason why gluons reggeize but photons do not. 

Total cross section is related to the forward part of the {\it elastic}
scattering amplitude, so only the exchange of colour-singlet object
contributes to it. The dominant amplitude then comes from the exchange
of two interacting Reggeons, indicated by $\delta_2$
in Fig.~2(b), or two non-interacting Reggeons $\delta_1^2$. The 
result is of order $\alpha_s^2$, and it is the BFKL Pomeron \cite{BFKL},
which as mentioned before violates unitarity.
$s$-channel unitarity and the Froissart bound are restored  when
we incorporate the singlet part of $p$-Reggeon exchanges, 
with all $p\ge 3$
included. For virtual-photon proton total cross section, as measured
at HERA \cite{ZEUS}, this results in a shallower growth of 
total cross section with energy for a smaller virtuality $Q^2$
of the virtual photon, as shown in Fig.~3.

For details see Ref.~\cite{DKL}.

Finally one might ask why coherence
 should have anything to do with high-energy collisions.
After all, the centre of collision will be hotter than the centre of
a star (Fig.~2(a)), whereas Bose-Einstein coherence usually 
happens at low
temperature. The answer is that although the centre is hot, the 
peripheral regions are `cold', and that is sufficient to produce
factorization as indicated in Fig.~2(b).

\begin{figure}[b!] 
\centerline{\epsfig{file=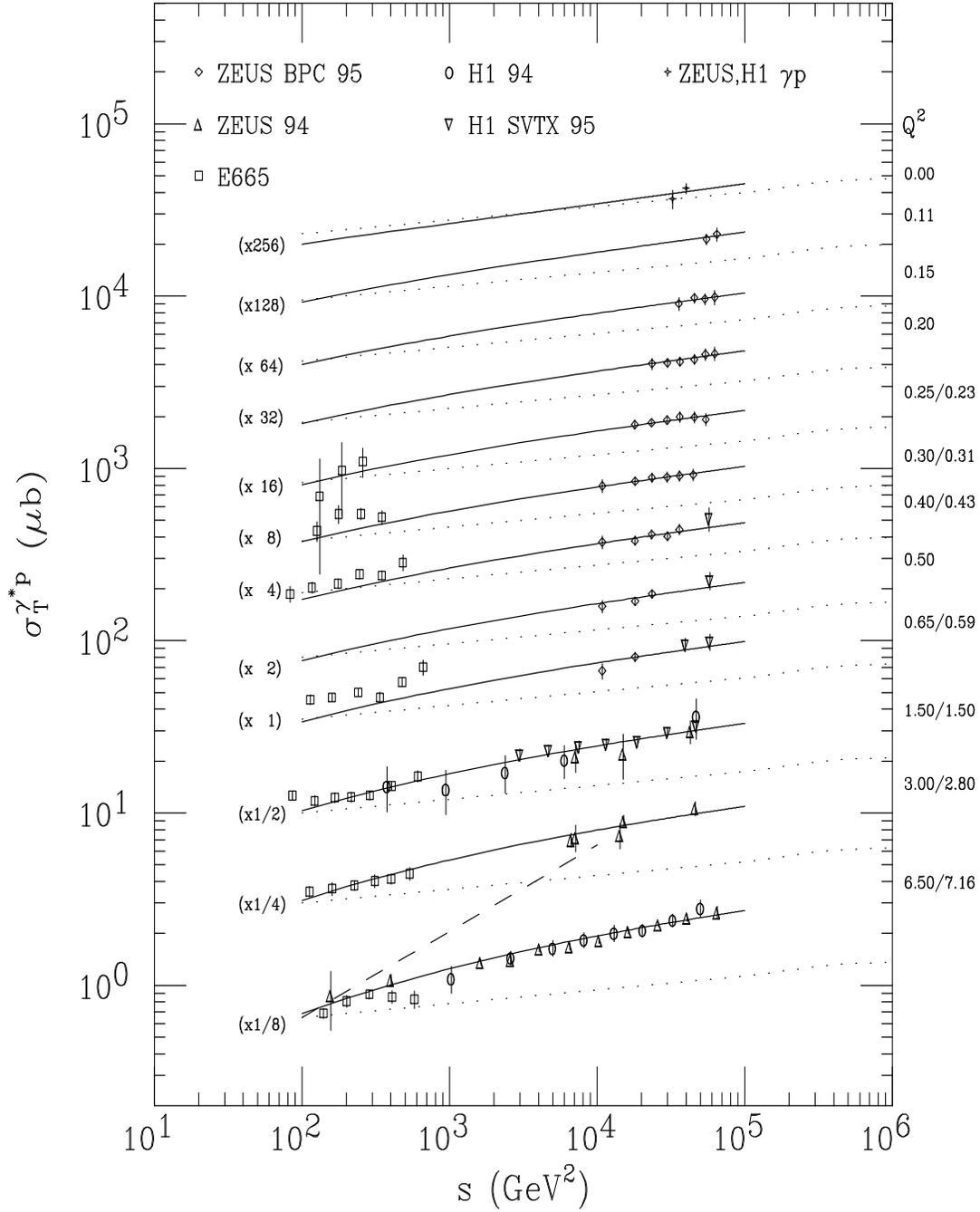,height=3.5in,width=3.5in}}
\vspace{10cm}
\caption{
Energy dependence of $\gamma^*p$ total cross section as a function
of photon virtuality $Q^2$. Data is from Ref.~[8]. The dotted
line represents a dependence of $s^{0.08}$ obeyed by all hadronic total
cross sections.
The dash line gives a $s^{0.5}$ variation predicted by
the leading-log BFKL Pomeron, and the solid line is the prediction of
the unitary theory in Ref.~[6].}
\label{fig3}
\end{figure}

\end{document}